\documentclass[aps,twocolumn,prd,showpacs,showkeys,preprintnumbers,superscriptaddress,nobibnotes,floatfix,longbibliography,nofootinbib]{revtex4-1}
\pdfoutput=1
\usepackage{amsmath}
\usepackage{amsfonts}
\usepackage{amssymb}
\usepackage{mathrsfs}
\usepackage{graphicx}
\usepackage{color}
\usepackage{longtable}
\usepackage{hyperref}
\usepackage{multirow}
\usepackage{slashed}
\usepackage{epsf}

\usepackage{color}

\begin{document}

\title{Right-Handed Slepton Bulk Region for Dark Matter in Generalized No-scale $\mathcal{F}$-$SU(5)$ with Effective Super-Natural Supersymmetry}

\author{Xiangwei Yin}
\email{yinxiangwei@itp.ac.cn}
\affiliation{CAS Key Laboratory of Theoretical Physics, Institute of Theoretical Physics,
	Chinese Academy of Sciences, Beijing 100190, China}
\affiliation{School of Physical Sciences, University of Chinese Academy of Sciences,
	No.~19A Yuquan Road, Beijing 100049, China}

\author{James A. Maxin}
\email{hep-cosmology@pm.me}
\affiliation{Department of Chemistry and Physics, Louisiana State University, Shreveport, LA 71115, USA}

\author{Dimitri V. Nanopoulos}
\email{dimitri@physics.tamu.edu}
\affiliation{George P. and Cynthia W. Mitchell Institute for Fundamental Physics and Astronomy,
	Texas A$\&$M University, College Station, TX 77843, USA}
\affiliation{Astroparticle Physics Group, Houston Advanced Research Center (HARC), Mitchell Campus, Woodlands, TX 77381, USA}
\affiliation{Academy of Athens, Division of Natural Sciences, 28 Panepistimiou Avenue, Athens 10679, Greece.}

\author{Tianjun Li}
\email{tli@itp.ac.cn}
\affiliation{CAS Key Laboratory of Theoretical Physics, Institute of Theoretical Physics,
	Chinese Academy of Sciences, Beijing 100190, China}
\affiliation{School of Physical Sciences, University of Chinese Academy of Sciences, No.~19A Yuquan Road, Beijing 100049, China}



\begin{abstract}
We propose Generalized No-Scale Supergravity, the simplest scenario for Effective Super-Natural Supersymmetry, naturally solving the supersymmetry electroweak fine-tuning problem and including natural dark matter. A light right-handed slepton bulk region is realized in $\mathcal{F}$-$SU(5)$ and the pMSSM. The bulk may be beyond the LHC reach, though can be probed at the 1000-day LUX-ZEPLIN, Future Circular Collider at CERN, Circular Electron Positron Collider, and Hyper-Kamiokande.
\end{abstract}
\maketitle


\textbf{Introduction~--}~A natural solution to the gauge hierarchy problem in the Standard Model (SM) is Supersymmetry (SUSY).  A few salient solutions to unexplained phenomena provided by supersymmetric SMs (SSMs) with R-parity are: (i) gauge coupling unification~\cite{gaugeunification}; (ii) Lightest Supersymmetric Particle (LSP) neutralino as a dark matter (DM) candidate~\cite{Goldberg:1983nd}; and (iii) electroweak  (EW)  gauge  symmetry can be broken radiatively  due  to  the  large  top quark Yukawa coupling. This list is by no means exhaustive,  but the first solution above deserves special emphasis, given that gauge coupling unification strongly implies Grand Unified Theories (GUTs)~\cite{Georgi:1974sy,Pati:1974yy,Mohapatra:1974hk,Fritzsch:1974nn,Georgi:1974my}, which might be constructed from superstring theory. These triumphs are evidence that SUSY builds a bridge between low-energy phenomenology and high-energy fundamental physics, leading to promising new physics beyond the SM.

As of the present, LHC SUSY searches have been barren of tantalizing signals of new physics, establishing strong constraints on the SSMs. Devoid of observation, SUSY lingers as only a beautiful yet unsubstantiated field theory. The prior ten years of proton-proton collisions without any verifiable ``bumps'' beyond SM expectations have elevated low mass bounds on the gluino, first-two generation squarks, stop, and sbottom to around 2.3~TeV, 1.9~TeV, 1.25~TeV, and 1.5~TeV, respectively~\cite{ATLAS-SUSY-Search, Aad:2020sgw, Aad:2019pfy, CMS-SUSY-Search-I, CMS-SUSY-Search-II}. Given these larger than anticipated lower limits on SUSY masses, we must unfortunately face the prospect that we may be encountering the SUSY EW fine-tuning (EWFT) problem. In that event, some encouraging and successful solutions to the EWFT problem have been proposed in the literature~\cite{Dimopoulos:1995mi, Cohen:1996vb, Kitano:2005wc, Kitano:2006gv, LeCompte:2011cn, LeCompte:2011fh, Fan:2011yu, Kribs:2012gx, Baer:2012mv, Baer:2012cf, Drees:2015aeo,Ding:2015epa,Baer:2015rja,Batell:2015fma,Fan:2014axa}.  In particular, we have proposed Super-Natural SUSY~\cite{Leggett:2014hha,Du:2015una, Li:2015dil} by considering No-Scale Supergavity (SUGRA)~\cite{Cremmer:1983bf}
and SUSY breaking soft terms in M-theory on $S^1/Z_2$\cite{Li:1998rn}, and demonstrated that the high-energy fine-tuning measure defined by Ellis-Enqvist-Nanopoulos-Zwirner~\cite{Ellis:1986yg} and Barbieri-Giudice~\cite{Barbieri:1987fn} (EENZ-BG) is of order one naturally, even if the supersymmetric particle (sparticle) spectra are heavy. Moreover, Super-Natural SUSY can be generalized to Effective Super-Natural SUSY~\cite{Ding:2015epa}.

Deriving an explanation of the observed dark matter relic density for a Bino dominant LSP remains another formidable issue. The methodology to resolve this dilemma typically comes in four distinct approaches: (1) Bulk region where the sfermions (supersymmetric partners of the SM fermions) are light; (2) The $Z$/Higgs funnel or $Z$/Higgs resonance, where the LSP neutralino mass is about half of the masses of the $Z$ boson, SM Higgs, $CP$-even Higgs $H_{0}$, or $CP$-odd Higgs $A_{0}$; (3) Coannihilation, where the sfermion masses are close to the LSP neutralino; or (4)  Mixing scenario or well-tempered scenario, where the LSP neutralino has enough Wino or Higgsino component to significantly increase the annihilation cross section. Evaluating these four scenarios, it seems to us that the bulk region may be the most natural. Demanding naturalness in both SUSY and dark matter is therefore a prominent pressing challenge, leading to the compelling question: Is it possible to have a viable bulk region for dark matter in a natural SUSY scenario?

In this paper, we consider ${\cal F}$-$SU(5)$, {\it i.e.}, the flipped $SU(5)\times U(1)_X$ GUT model with extra TeV-scale vector-like particles~\cite{Jiang:2006hf} that have been constructed systematically in local F-theory model building~\cite{Jiang:2009zza, Jiang:2009za}. Alternatively, these models can also be realized in free fermionic string constructions~\cite{Lopez:1992kg}. Super-Natural SUSY via No-Scale SUGRA is not a resolution to the SUSY EWFT problem in the specific case when there is also a light Bino LSP, due to a correlation of the Bino mass with the Wino and gluino masses. Therefore, we propose Generalized No-Scale SUGRA, where Effective Super-Natural SUSY can be realized. In order to uncover the bulk region for dark matter, we can only consider light right-handed sleptons given that the LHC SUSY searches indicate that all other sfermions must be heavy. First, a determination must be carried out as to whether an interaction between sfermions and the LSP is coannihilation or annihilation. Rendering a judgement involves inspecting the mass difference between the light right-handed sfermions and LSP, though we mention that the ratio of the mass difference $\mathcal{R}_{\phi}\equiv (m_{\phi}-m_{\Tilde{\chi}_{1}^{0}})/m_{\Tilde{\chi}_{1}^{0}}$ is more important than the absolute mass difference, where $\phi$ is $\Tilde{\tau}_{1}$ (light stau) or $\Tilde{e}_R$ (light selectron). Comprehensive numerical studies that we present in this work show that $\mathcal{R}_{\phi} \gtrsim 10\%$ is a conservative criterion to formulate the bulk region, {\it i.e.}, the observed dark matter density is obtained via traditional annihilations, not from coannihilations or resonances, etc. We also investigate outside the tight ${\cal F}$-$SU(5)$ constraints by evaluating the phenomenological Minimal Supersymmteric Standard Model (pMSSM).

{\bf The $\mathcal{F}$-$SU(5)$ Model~--}~For a thorough review of ${\cal F}$-$SU(5)$, we invite the reader to explore the wealth of literature on the model~\cite{Li:2010ws,Li:2011ab,Li:2021cte}, and references therein, including the No-Scale flipped $SU(5)$ string derived inflationary model, cosmology, and gravitational waves~\cite{Ellis:2019jha}. Here we only illuminate a few of the most vital attributes. In the flipped $SU(5)$ models~\cite{F-SU5}, there are three families of SM fermions whose quantum numbers under the $SU(5)\times U(1)_{X}$ gauge group are
$F_i={\mathbf{(10, 1)}},~ {\bar f}_i={\mathbf{(\bar 5, -3)}},~{\bar l}_i={\mathbf{(1, 5)}}$, where $i=1, 2, 3$. To break the GUT and electroweak gauge symmetries, we introduce two pairs of Higgs fields $H={\mathbf{(10, 1)}},~{\overline{H}}={\mathbf{({\overline{10}}, -1)}}, ~h={\mathbf{(5, -2)}},~{\overline h}={\mathbf{({\bar {5}}, 2)}}$. True string-scale gauge coupling unification~\cite{Jiang:2006hf, Jiang:2009zza, Jiang:2009zza} is achieved by introduction of the vector-like particles (dubbed ``flippons'') $XF={\mathbf{(10, 1)}}~,~{\overline{XF}}={\mathbf{(\overline{10}, 1)}}~,~Xl={\mathbf{(1, -5)}}~,~{\overline{Xl}}={\mathbf{(1, 5)}}$, which form complete flipped $SU(5)\times U(1)_X$ multiplets. The flipped $SU(5)$ model with the additional vector-like particles derived from local F-Theory model building is referred to as $\mathcal{F}$-$SU(5)$. The Renormalization Group Equation (RGE) $\beta$ coefficients undergo a shift due to the vector-like multiplets, lifting the $SU(5)\times U(1)_X$ unification to the string scale around $10^{17}$~GeV (correlating to the mass scales $M_5$ and $M_{1X}$), adjacent to the Planck scale. Subsequently, a second stage unification $SU(3)_C\times SU(2)_L$ (defined as mass scale $M_{32}$) occurs near the traditional MSSM GUT scale around $10^{16}$~GeV. The separated unification structure produces a flat $SU(3)$ RGE running (due to a vanishing $b3$ coefficient enforced by the vector-like multiplets) from the GUT scale to the TeV scale where all vector-like multiplets decouple, defined as the scale $M_V$. The universal vector-like particle mass scale $M_V$ is treated as a free model parameter. A light $M_V$ mass scale ($M_V \lesssim 10$~TeV) allows for a larger vector-like particle Yukawa coupling, contributing to the light Higgs boson mass~\cite{Li:2011ab}.

{\bf Generalized No-Scale SUGRA~--}~In order to generate a light Bino, evade the large LHC constraint on the gluino mass, and sustain naturalness conditions, the generalization of No-Scale SUGRA is essential. At the $SU(5)\times U(1)_X$ unification scale (string scale), we vary the $SU(5)$ gaugino mass $M_5$ from 1200~GeV to 5000~GeV, yielding a large gluino mass. To produce a light Bino, we vary the $U(1)_X$ gaugino mass $M_{1X}$ from 100~GeV to 600~GeV. Note that No-Scale SUGRA is obtained at tree level and can be violated at one loop, so we assume the universal supersymmetry breaking soft mass $M_0$ and trilinear soft term $A_0$ are smaller than about 1\% of $M_5$.
Finally, we span $\tan\beta$ from 2 to 65, and the vector-like particle mass scale $M_V$ from 1 TeV to 10 TeV.

According to Effective Super-Natural SUSY~\cite{Ding:2015epa}, we have shown that a supersymmetry breaking scenario is natural 
if all the fundamental parameters that have large EENZ-BG fine-tuning measures are correlated. In our generalized No-Scale SUGRA presented here, 
the fine-tuning measures for the SUSY breaking soft terms $M_{1X}$, $M_0$, and $A_0$ are all small, and only $M_5$ might have a large
fine-tuning measure. Therefore, our generalized No-Scale SUGRA is approximately Super-Natural SUSY, and thus indeed natural. More specifically, it is only a small deviation from Super-Natural SUSY, and hence the simplest scenario for Effective Super-Natural SUSY, where only one fundamental parameter may have a large EENZ-BG fine-tuning measure.

{\bf Numerical Procedure~--}~To investigate the bulk region in Generalized No-Scale $\mathcal{F}$-$SU(5)$ and target the requisite $M_5$ and $M_{1X}$ gaugino mass scales, we need to focus on small $M_{1}$. Accomplishing this goal requires we exploit the relationship in $\mathcal{F}$-$SU(5)$ between $M_{1}$ for $U(1)_{Y}$ and $M_{1X}$ for $U(1)_{X}$ at the $M_{32}$ scale, which is~\cite{Li:2010rz}:
\begin{equation}
\frac{M_1}{\alpha_1} \equiv \frac{24}{25} \frac{M_{1 X}}{\alpha_{1 X}}+\frac{1}{25} \frac{M_5}{\alpha_5}~,
\end{equation}
where $\alpha_i$ are the gauge couplings at their respective scales. Conjointly, the following experimental constraints are imposed:
\begin{itemize}
    \item Require neutralino LSP
    \item Constraints on the mass of gluino and first and second generation squark masses~\cite{ATLAS:2017mjy,Vami:2019slp,CMS:2017okm} of 
    $m_{\Tilde{g}}\gtrsim 2.2 ~\text{TeV} $,~$m_{\Tilde{q}}\gtrsim 2.0 ~\text{TeV}$
    \item Rare B-meson decay constraint of~\cite{CMS:2014xfa} $1.6 \times 10^{-9}$ $\leq$ BR($B_{s}^0 \rightarrow \mu^{+}\mu^{-}$) $\leq$ $4.2 \times 10^{-9}$ and branching ratio of rare b-quark decay of $2.99 \times 10^{-4} \leq$ BR($b\rightarrow s \gamma$) $\leq 3.87 \times 10^{-4}$~\cite{HFLAV:2014fzu}
    \item Attention to both the experimental measurement of the light Higgs boson mass and its theoretical uncertainty~\cite{Slavich:2020zjv,Allanach:2004rh}, applying a range 122 GeV$\leq m_{h} \leq$ 128 GeV
    \item Constraints on spin-independent DM-nuclei cross sections from XENONnT~\cite{XENON:2023cxc} and LUX-ZEPLIN~\cite{LZ:2018qzl,LZ:2022lsv}
    \item Relic density of cold DM measured by the 5$\sigma$ Planck 2018~\cite{Planck:2018nkj} of $0.114 \leq \Omega_{\rm DM}h^{2} \leq 0.126$ where below this range is regarded as under-saturated and above is over-saturated.
\end{itemize}

All SUSY RGE numerical calculations are executed with MicrOMEGAs 2.1~\cite{Belanger:2001fz} incorporating a proprietary revision to the SuSpect 2.34~\cite{Djouadi:2002ze} codebase to evolve vector-like particle and $\mathcal{F}$-$SU(5)$ enhanced RGEs. We deploy MicrOMEGAs to compute the relic density in the instances of both annihilation and coannihilation, plus, to ensure we are wielding the correct number for annihilation only, a parallel computation valid when coannihilation is absent is performed, given by~\cite{Arkani-Hamed:2006wnf}
\begin{equation}
\footnotesize{
\Omega_{\tilde{B}} h^2=1.3 \times 10^{-2}\left(\frac{m_{\tilde{e}_R}}{100 \mathrm{GeV}}\right)^2 \frac{(1+r)^4}{r\left(1+r^2\right)}\left(1+0.07 \log \frac{\sqrt{r} 100 \mathrm{GeV}}{m_{\tilde{e}_R}}\right)
\label{relic}
}
\end{equation}
where $m_{\tilde{e}_R}$ the mass of right-handed sleptons, $r \equiv  M_1^2 / m_{\tilde{e}_R}^2$, and $M_1$ is the Bino mass.

Investigation of the bulk region for a pure Bino-like neutralino necessitates avoiding resonance annihilation and coannihilation. Prevention demands the following conditions be implemented:
\begin{itemize}
    \item 99.9\% Bino-like LSPs are selected to prohibit large annihilation cross sections induced by Higgsino or Wino components
    \item $2 m_{\Tilde{\chi}_{1}^{0}} \ll m_{H^{0}}$,$m_{A^{0}}$ and $2 m_{\Tilde{\chi}_{1}^{0}} \gg m_{h}$ are enforced to avoid the "Higgs funnel"
    \item First and second right-handed sleptons, $\Tilde{\tau}_{1}$ and $\Tilde{e}_{R}$, are naturally light, so coannihilation processes are negligible when $\mathcal{R}_{\Tilde{\tau}_{1}}\equiv \frac{m_{\Tilde{\tau}_{1}}-m_{\Tilde{\chi}_{1}^{0}}}{m_{\Tilde{\chi}_{1}^{0}}} \gtrsim 10 \%$ and $\mathcal{R}_{\Tilde{e}_{R}}\equiv \frac{m_{\Tilde{e}_{R}}-m_{\Tilde{\chi}_{1}^{0}}}{m_{\Tilde{\chi}_{1}^{0}}} \gtrsim 10 \%$
\item Require the SM Higgs resonance to vanish, which transpires when $|\mu|^{2} \gg M_{Z}^{2}$, via the coupling $g_{h{\Tilde{\chi}_{1}^{0}}{\Tilde{\chi}_{1}^{0}}} \propto \frac{M_{Z}(2\mu \text{cos}\beta+ M_{1})}{\mu^{2}-M_{1}^{2}}$~\cite{Djouadi:2005dz}.
\end{itemize}

{\bf Numerical Results~--}~Constructing the bulk region requires we consider only traditional annihilations. In order to assure that our calculation using only annihilations is accurate, a consistency check was performed between two disparate methods, the relic density calculated by micrOMEGAs and the formula in Eq.(\ref{relic}). We found the two distinct computations were consistent. With the validity of our annihilation calculation method now established, the expression $\mathcal{R}_{\phi} < 10\%$ is tested to authenticate that it does in fact deliver a large coannihilation percentage. Indeed, for $\mathcal{R}_{\phi} = 5 - 8\%$, the relic density computed using only annihilations deviates from the coannihilation computation by $\geq 50\%$. The large deviation between the two results verifies the use of $\mathcal{R}_{\phi} < 10\%$ for identifying a large coannihilation factor. On the other hand, numerical results reveal that for $\mathcal{R}_{\phi} = 10 - 12\%$, the annihilation and coannihilation calculations only deviate by $20 - 30\%$, a negligible amount of coannihilation compared to $70 - 80\%$ annihilation.

\begin{figure}[tp]
    \centering
    \includegraphics[width=0.45\textwidth]{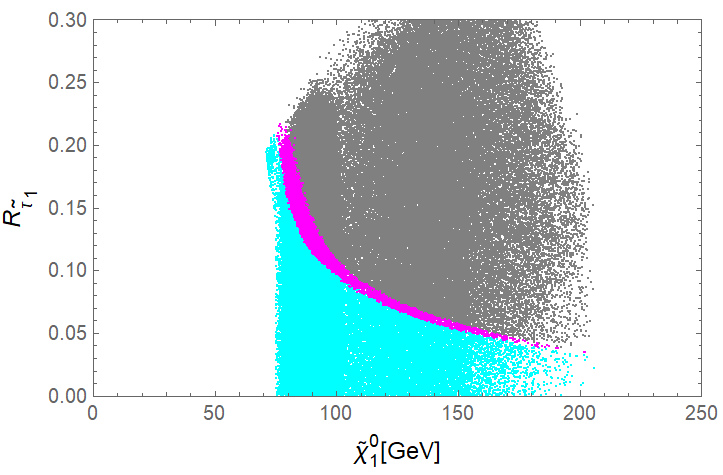}
    \caption{Bulk region in Generalized No-Scale $\mathcal{F}$-$SU(5)$. Cyan, magenta, gray points correspond to under-saturated, saturated, over-saturated DM relic density.}
    \label{FSU5-stau1}
\end{figure}

\begin{figure}[tp]
    \centering
    \includegraphics[width=0.45\textwidth]{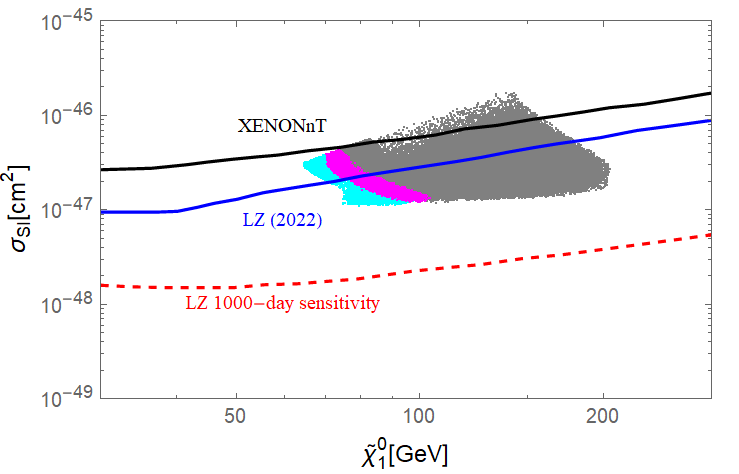}
    \caption{Generalized No-Scale $\mathcal{F}$-$SU(5)$ bulk region LSPs plot in reference to spin-independent DM-nuclei cross sections from XENONnT~\cite{XENON:2023cxc} and LUX-ZEPLIN\cite{LZ:2018qzl,LZ:2022lsv}. Cyan, magenta, gray points correspond to under-saturated, saturated, over-saturated DM relic density with $\mathcal{R}_{\Tilde{\tau}_{1}} \gtrsim 10\%.$ We underscore the significance of the 1000-day LUX-ZEPLIN run that should fully probe the $\mathcal{F}$-$SU(5)$ bulk and about 50\% of the pMSSM bulk (not shown).}
    \label{LZ1000}
\end{figure}

\begin{figure}[tp]
    \centering
    \includegraphics[width=0.45\textwidth]{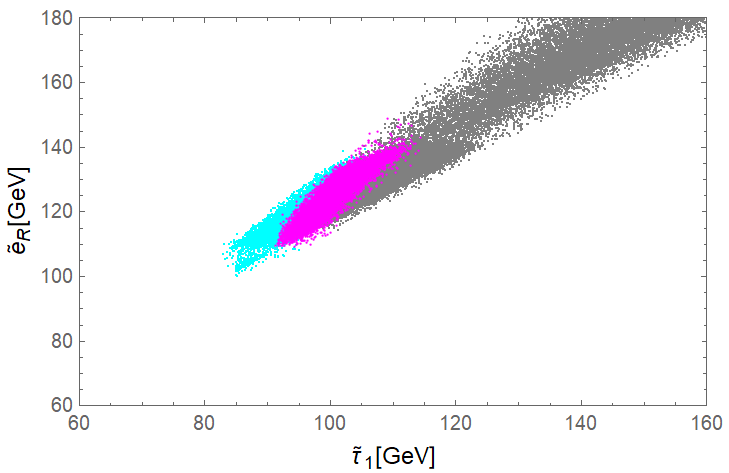}
    \caption{Light right-handed slepton masses in the Generalized No-Scale $\mathcal{F}$-$SU(5)$ bulk region. Cyan, magenta, gray points correspond to under-saturated, saturated, over-saturated DM relic density with $\mathcal{R}_{\Tilde{\tau}_{1}} \gtrsim 10\%.$ }
    \label{FSU5slepton}
\end{figure}

The bulk region in Generalized No-Scale $\mathcal{F}$-$SU(5)$ is illustrated in FIG.~\ref{FSU5-stau1}, with $\mathcal{R}_{\Tilde{\tau}_{1}}$ plot as a function of the Bino-like neutralino $m_{\Tilde{\chi}_{1}^{0}}$. The mass hierarchy in $\mathcal{F}$-$SU(5)$ is $m_{\Tilde{\chi}_{1}^{0}}\textless m_{\Tilde{\tau}_{1}} \textless m_{\Tilde{e}_{R}} = m_{\Tilde{\mu}_{R}}$, hence, $\mathcal{R}_{\Tilde{e}_{R}}$ always exceeds $\mathcal{R}_{\Tilde{\tau}_{1}}$. All points in FIG.~\ref{FSU5-stau1} satisfy the itemized experimental constraints specified in prior section. Note that No-Scale $\mathcal{F}$-$SU(5)$ is in tension with the recent muon anomalous magnetic moment measurements~\cite{Muong-2:2023cdq}, though this can be fully remedied with the addition of a SM singlet and chirality flip~\cite{Li:2021cte}. Cyan, magenta, gray points in FIG.~\ref{FSU5-stau1} correspond to the under-saturated, saturated, and over-saturated DM relic density, respectively. If the Bino contributes all the DM abundance, the ratio $\mathcal{R}_{\Tilde{\tau}_{1}}\gtrsim 10\%$ implies $m_{\Tilde{\chi}_{1}^{0}}\leq 103.0$ GeV. The light Bino LSP of the bulk region could be fully probed within the next few years, where FIG.~\ref{LZ1000} shows the compelling expected sensitivity of the 1000-day LUX-ZEPLIN experiment~\cite{LZ:2018qzl} and broad coverage of the No-Scale $\mathcal{F}$-$SU(5)$ bulk region. The $\Tilde{\tau}_{1}$, $\Tilde{e}_{R}$ plane is delineated in FIG.~\ref{FSU5slepton}. Within the Generalized No-Scale $\mathcal{F}$-$SU(5)$ bulk region, FIG.~\ref{FSU5slepton} shows the upper limit of $\Tilde{\tau}_{1}$ and $\Tilde{e}_{R}$ are around 115 GeV and 150 GeV, respectively. Recognize that these right-handed sleptons and Bino LSP are $naturally$ light, thus, the LSP has not been fine-tuned to fortuitously conform to Planck satellite $5 \sigma$ relic density observations. Two benchmark points are provided in TABLE~\ref{FSU5-T1} highlighting all the parameters and constraints discussed here. 

\begin{figure}[tp]
    \centering
    \includegraphics[width=0.45\textwidth]{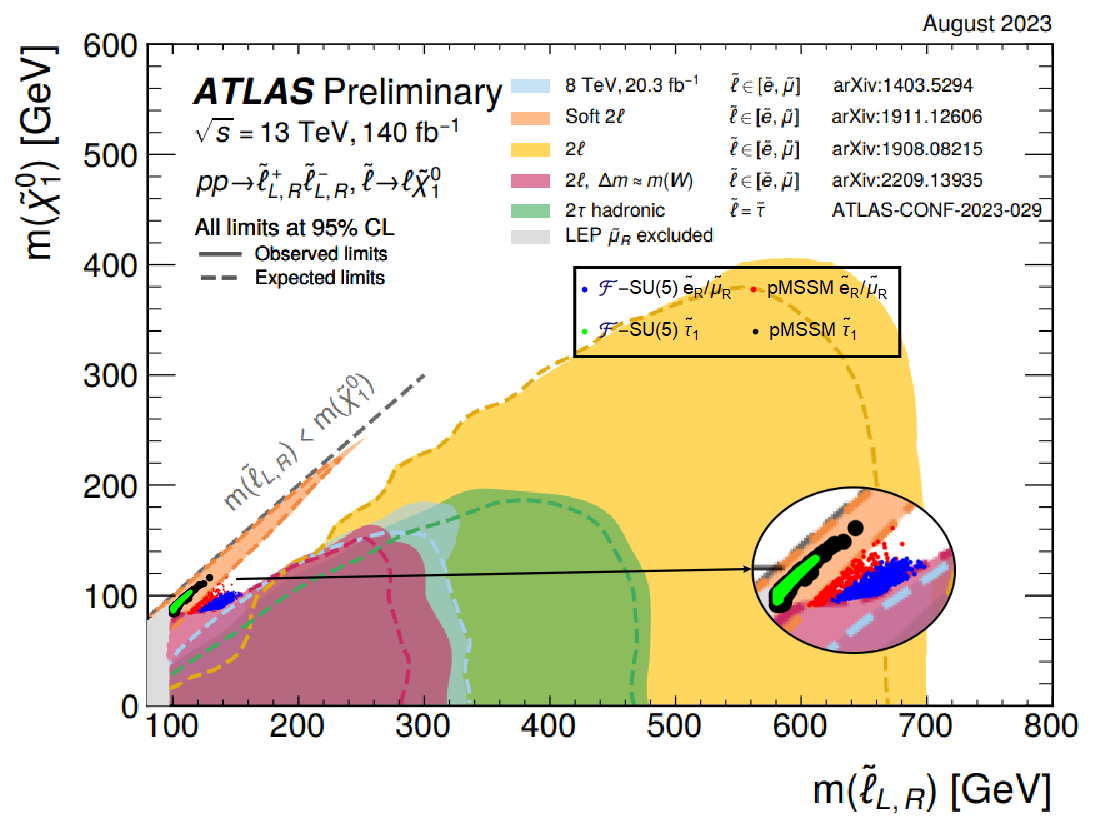}
    \caption{Generalized No-Scale $\mathcal{F}$-$SU(5)$ and pMSSM bulk regions superimposed over the August 2023 ATLAS Summary Plot~\cite{ATLAS:2023xco} of SUSY searches for electroweak production of sleptons~\cite{ATLAS:2014zve,ATLAS:2019lng,ATLAS:2019lff,ATLAS:2022hbt,ATLAS:2023djh}. The blue points correspond to $\Tilde{e}_{R}$ = $\Tilde{\mu}_{R}$ in $\mathcal{F}$-$SU(5)$, whereas the red points are $\Tilde{e}_{R}$ = $\Tilde{\mu}_{R}$ in the pMSSM. The green points correspond to $\Tilde{\tau}_{1}$ in $\mathcal{F}$-$SU(5)$, while the black points are $\Tilde{\tau}_{1}$ in the pMSSM. The inset is a zoom of the bulk. All points plot here are annihilation only per our requirement $\mathcal{R}_{\phi} \gtrsim 10\%$. Note the ATLAS orange shaded sliver applies to $\Tilde{e}_{R}$ = $\Tilde{\mu}_{R}$ only, and not the $\Tilde{\tau}_{1}$. The $\Tilde{\tau}_{1}$ constraints are the ATLAS green shaded region, which both $\mathcal{F}$-$SU(5)$ and pMSSM points are comfortably beyond. The evasion of the bulk region at the LHC thus far is evident.}
    \label{ATLAS-FSUT}
\end{figure}

\begin{figure}[tp]
    \centering
    \includegraphics[width=0.45\textwidth]{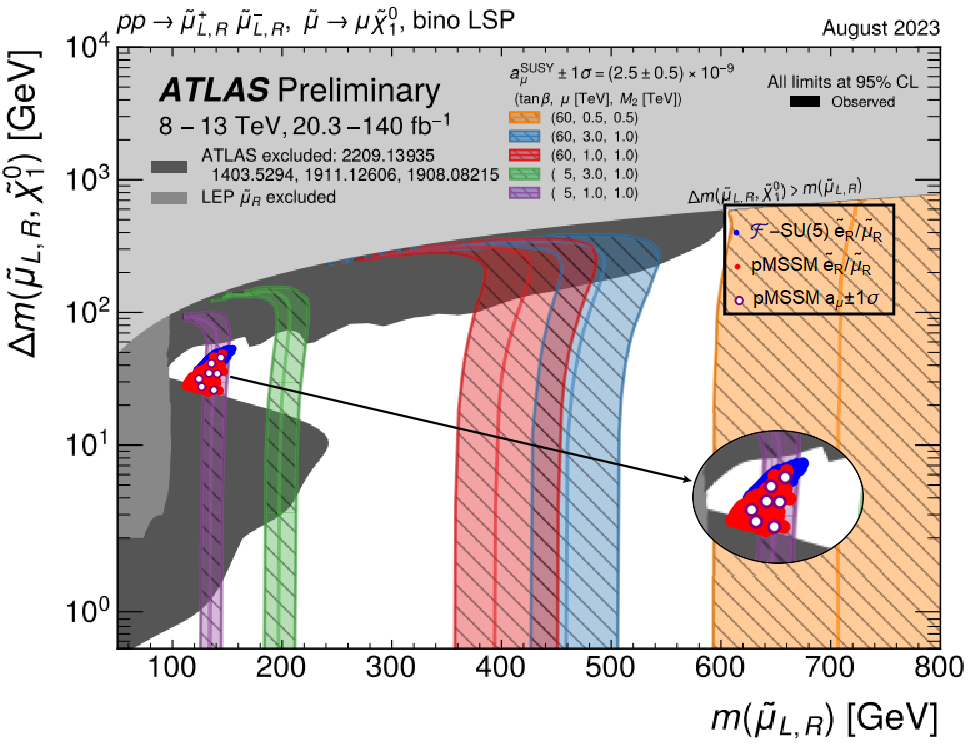}
    \caption{Generalized No-Scale $\mathcal{F}$-$SU(5)$ and pMSSM bulk regions superimposed over the August 2023 ATLAS Summary Plot~\cite{ATLAS:2023xco} of SUSY searches for electroweak production of smuons~\cite{ATLAS:2014zve,ATLAS:2019lng,ATLAS:2019lff,ATLAS:2022hbt}, plot here in terms of $\Delta m = (\tilde{\mu}_{L,R}, \tilde{\chi}_1^0)$ for a Bino LSP, emphasizing consistency of the bulk with recent muon anomalous magnetic moment measurements~\cite{Muong-2:2023cdq}. The inset is a zoom of the bulk.}
    \label{DeltaM}
\end{figure}

The August 2023 ATLAS Summary Plot~\cite{ATLAS:2023xco} highlighting observed exclusion limits from ATLAS SUSY searches for electroweak production of sleptons~\cite{ATLAS:2014zve,ATLAS:2019lng,ATLAS:2019lff,ATLAS:2022hbt,ATLAS:2023djh} is shown in FIG.~\ref{ATLAS-FSUT} with several Generalized No-Scale $\mathcal{F}$-$SU(5)$ benchmark points in the bulk region superimposed. Though not shown in this paper, the situation is similar with regard to CMS SUSY searches for electroweak production of sleptons~\cite{CMS:2020bfa,CMS:2023qhl,CMS:2022syk}. All points plot in FIG.~\ref{ATLAS-FSUT} are primarily traditional annihilation only and adhere to our requirement $\mathcal{R}_{\phi} \gtrsim 10\%$. The blue points correspond to $\Tilde{e}_{R}$ = $\Tilde{\mu}_{R}$ in $\mathcal{F}$-$SU(5)$, whereas the green points correspond to $\Tilde{\tau}_{1}$ in $\mathcal{F}$-$SU(5)$. It is important to point out that the ATLAS thin orange shaded region in FIG.~\ref{ATLAS-FSUT} applies only to $\Tilde{e}_{R}$ = $\Tilde{\mu}_{R}$, and not the $\Tilde{\tau}_{1}$. The ATLAS green shaded region in FIG.~\ref{ATLAS-FSUT} depicts the $\Tilde{\tau}_{1}$ constraints, which $\mathcal{F}$-$SU(5)$ is well beyond. A corollary picture of the bulk at the LHC is depicted in FIG.~\ref{DeltaM}, with $\mathcal{F}$-$SU(5)$ points superimposed over an August 2023 ATLAS Summary plot for smuon SUSY searches~\cite{ATLAS:2014zve,ATLAS:2019lng,ATLAS:2019lff,ATLAS:2022hbt}. In this particular instance the smuon mass is plot as a function of $\Delta m = (\tilde{\mu}_{L,R}, \tilde{\chi}_1^0)$ for a Bino LSP in order to emphasize probing of those regions consistent with recent muon anomalous magnetic moment measurements~\cite{Muong-2:2023cdq}. The predicament depicted in FIGs.~\ref{ATLAS-FSUT}~-~\ref{DeltaM} is that given the compressed nature of these spectra, this bulk region may not be probed at the LHC, though these light sleptons could conceivably be observed when the forthcoming circular colliders power up their beams, namely the Future Circular Collider (FCC-ee)~\cite{FCC:2018byv,FCC:2018evy} at CERN and the Circular Electron-Positron Collider (CEPC)~\cite{CEPCStudyGroup:2018ghi} with its sensitivity specified in Ref.~\cite{Yuan:2022ykg}. Likewise, the proton lifetime via dimension-six proton decay is near $3-4 \times 10^{34}$~years, so this ``fast'' dimension-six proton decay is within reach of the future Hyper-Kamiokande experiment~\cite{Hyper-Kamiokande:2018ofw}. 

The methodology just discussed is extended to include the much less constrained generic pMSSM. The pMSSM contains 22 free parameters, and we input $M_{A}$ and $\mu$ in lieu of $m_{H_{u}}^{2}$ and $m_{H_{d}}^{2}$. The scanning ranges of the pMSSM parameters are as follows:

\begin{equation}
\footnotesize{
\begin{split}
   &20 ~\text{GeV}\leq M_{1}\leq 1000 ~\text{GeV} \qquad\quad 2\leq tan\beta\leq 65 \\
   &1000 ~\text{GeV}\leq M_{2}\leq 5000 ~\text{GeV} \qquad 1000 ~\text{GeV}\leq M_{A}, \mu \leq 6000 ~\text{GeV}\\
   &1200 ~\text{GeV}\leq M_{3}\leq 5000 ~\text{GeV}   \qquad   M_{1} \leq m_{\Tilde{e}_{R}}, m_{\Tilde{\tau}_{R}}      \leq 2 M_{1} \\
   &2500 ~\text{GeV}\leq m_{\Tilde{q}},m_{\Tilde{Q}},m_{\Tilde{u}_{R}},m_{\Tilde{t}_{R}},m_{\Tilde{d}_{R}},m_{\Tilde{b}_{R}}    \leq 5000 ~\text{GeV} \\
   &700 ~\text{GeV}\leq m_{\Tilde{l}}\leq 2000 ~\text{GeV}\qquad \quad 1200 ~\text{GeV}\leq m_{\Tilde{L}}\leq 5000 ~\text{GeV} \\
  -&5000 ~\text{GeV}\leq A_{u}, A_{d}, A_{e}, A_{t}, A_{b}, A_{\tau} \leq 5000 ~\text{GeV}
\end{split}
}
\end{equation}

\begin{table}[ht]
\begin{center}
\begin{tabular}{cccc}
     \hline \hline
     $M_{5}$ &3996 & 2591 \\
     $M_{1X}$&473 &268\\
     $M_{0}$ & 23.15& 21.49 \\
     $A_{0}$ & 0&0 \\
     $tan\beta$&3.04 & 2.53 \\
     $M_{V}$&9063 & 4603 \\
     \hline \hline
     $m_{h}$& 125.43&123.06  \\
     $m_{A}$& 7325& 5395 \\
     \hline
     $m_{\Tilde{\chi}_{1}^{0}}$& 161.3&92.3 \\
     $m_{\Tilde{\tau}_{1}}$&169.6&103.4 \\
     $m_{\Tilde{e}_{R}}$&216.1 &130.1 \\
     $m_{\Tilde{t}_{1}}$&4273 &2747\\
     $m_{\Tilde{g}}$&4986 & 3259 \\
     $m_{\Tilde{u}_{R}}$&6798 &4606 \\
     \hline
     $\text{BR}(B_S^0 \rightarrow  \mu^+ \mu^-)\times10^{-9}$&3.03 & 3.05 \\
     $\text{BR}(b\rightarrow s \gamma)\times10^{-4}$& 3.61&3.61 \\
     $\sigma_{SI} \times 10^{-12} {\rm pb} $&6.28 & 17.10 \\
     $\tau_{p}\times 10^{34} {\rm yrs}$&5.01 & 3.95 \\
     \hline
     $\mathcal{R}_{\Tilde{\tau}_{1}}$&$5\%$&$12\%$&\\
     $\Omega_{\Tilde{\chi}}h^{2}$&0.1256 & 0.118 \\
     $\Omega_{\Tilde{\chi}}h^{2}$(No co-annihilation)& 0.386 &  0.147\\
     co-annihilation rate&$> 50\%$&$\sim 20\%$\\
     \hline
\end{tabular}
\end{center}
    \caption{Two benchmark points for Generalized No-Scale $\mathcal{F}$-$SU(5)$. All masses are in GeV.}
     \label{FSU5-T1}
\end{table}

\noindent where $m_{\Tilde{q}}$, $m_{\Tilde{u}_{R}}$, $m_{\Tilde{d}_{R}}$, $m_{\Tilde{l}}$, and $m_{\Tilde{e}_{R}}$ are the first/second generation sfermion mass parameters and $m_{\Tilde{Q}}$, $m_{\Tilde{t}_{R}}$, $m_{\Tilde{b}_{R}}$, $m_{\Tilde{L}}$, and $m_{\Tilde{\tau}_{R}}$ are third generation sfermions. In contrast with $\mathcal{F}$-$SU(5)$, there is no rigid mass ordering amongst $\Tilde{e}_{R}$ and $\Tilde{\tau}_{1}$ in the pMSSM. Our search imposes $\mathcal{R}_{\Tilde{e}_{R}} > \mathcal{R}_{\Tilde{\tau}_{1}}$ while floating the ratio $\mathcal{R}_{\Tilde{\tau}_{1}}$, and vice versa.  Numerical findings disclose the ratio $\mathcal{R}_{\Tilde{\tau}_{1}}\gtrsim10\%$ implies $m_{\Tilde{\chi}_{1}^{0}} \leq 117.7~\text{GeV}$. In the latter case where we imposed $\mathcal{R}_{\Tilde{\tau}_{1}} > \mathcal{R}_{\Tilde{e}_{R}}$, all pMSSM points with an $\Tilde{e}_{R}$ next to lightest supersymmetric particle (NLSP) are excluded by the ATLAS soft lepton SUSY search~\cite{ATLAS:2019lng}. Therefore, like Generalized No-Scale $\mathcal{F}$-$SU(5)$, the only viable pMSSM region in the bulk is for the case $m_{\Tilde{\chi}_{1}^{0}}\textless m_{\Tilde{\tau}_{1}} \textless m_{\Tilde{e}_{R}} = m_{\Tilde{\mu}_{R}}$. We only scanned for small $\tilde{e}_R$ in the pMSSM, but in principle $\tilde{e}_R$ can be much heavier than $\tilde{\tau}_1$ in the pMSSM. The pMSSM bulk is not included in FIG.~\ref{LZ1000}, though the 1000-day LUX-ZEPLIN experiment~\cite{LZ:2018qzl} is anticipated to probe about 50\% of the pMSSM bulk. The ATLAS SUSY search exhibited in FIG.~\ref{ATLAS-FSUT} also superimposes points from the bulk region in the pMSSM. The red points are $\Tilde{e}_{R}$ = $\Tilde{\mu}_{R}$ in the pMSSM, while the black points are $\Tilde{\tau}_{1}$ in the pMSSM. In the pMSSM, the bulk alone can explain recent muon anomalous magnetic moment measurements~\cite{Muong-2:2023cdq}. Indeed, pMSSM bulk region points are plot in FIG.~\ref{DeltaM} along with $\mathcal{F}$-$SU(5)$, and consistency with the anomalous magnetic moment results for small tan$\beta$ is clear (ATLAS generated purple band). The conclusions leaping from FIGs.~\ref{ATLAS-FSUT}~-~\ref{DeltaM} for both $\mathcal{F}$-$SU(5)$ and the pMSSM are uniform; the bulk region has so far stealthily eluded the reach of the LHC, potentially leaving the prospect of discovery residing with dark matter direct detection experiments and next generation circular colliders.

{\bf Conclusion~--}~The void of a substantiated SUSY signal at the LHC has promoted naturalness to an indispensible foundation of SUSY GUT models. It is not sufficient for a model to only attack and then solve the electroweak fine-tuning problem (EWFT), the model should also accomodate natural dark matter. The challenge resounds as to whether such broad naturalness is even posssible to formulate, and if so, could such a sweeping natural model even be probed at the LHC? 

The EWFT problem can be elegantly solved with Effective Super-Natural SUSY, but this technique alone cannot naturally generate light sfermions and thus a light Bino LSP, a scenario we regard as the most natural dark matter. In pursuit of the ambitious objective to construct a fully natural model, we proposed in this work a new perspective on No-Scale Supergravity (SUGRA), defining it Generalized No-Scale SUGRA. Arising from our generalization of No-Scale SUGRA is a marriage of natural SUSY with natural dark matter, a very favorable and welcome merger. The blissful union of natural SUSY with natural dark matter was fullfilled in the GUT model $\mathcal{F}$-$SU(5)$, deriving a region of the model space that naturally supports light right-handed sleptons and a light LSP, known as the bulk region, where $m_{\Tilde{\chi}_{1}^{0}}\leq 103.0$ GeV with negligible coannililation and upper limits on $m_{\Tilde{\tau}_{1}}$ and $m_{\Tilde{e}_{R}}$ about 115 GeV and 150 GeV, respectively. The $\mathcal{F}$-$SU(5)$ bulk region LSPs will receive full coverage during the presently running 1000-day LUX-ZEPLIN experiment. Our analytical results unveiled that the light right-handed sleptons in the bulk could be beyond the LHC reach, though may be probed amidst the imminent era of advanced circular colliders, for example, the Future Circular Collider (FCC-ee) at CERN and Circular Electron Positron Collider (CEPC). Equally meaningful, the fast proton decay could be observed at the future Hyper-Kamiokande experiment.

We extended the analysis beyond $\mathcal{F}$-$SU(5)$ and scrutinzed the pMSSM, exposing comparable results. We are partial to the beautiful simplicity of the stringy physical model No-Scale $\mathcal{F}$-$SU(5)$ and its unprecedented cosmology, yet we also recognize the advantages of studying the larger and much less constrained generic pMSSM. Our expanded investigation revealed a bulk region can subsist in the pMSSM also, though only with a light stau NLSP, where $m_{\Tilde{\chi}_{1}^{0}} \leq 117.7~\text{GeV}$ with negligible coannihilation. 

Should the LHC fail in the next few years to mount another substantial discovery beyond the light Higgs boson, a strategic plan forged by the high-energy physics community will undoubtably emerge. We suggest the tactics presented here that can formulate a model supporting both natural SUSY and natural dark matter be given explicit deliberation as to why discovery has been delayed and where to focus the forthcoming circular colliders.

\textbf{Acknowledgments~--}~We are indebted to Wenxing Zhang, Chuang Li, Xiaochuan Wang, and Jinmian Li for helpful discussions. We are also thankful to the ChinaHPC and HPC Cluster of ITP-CAS. This research is supported in part by the National Key Research and Development Program of China Grant No. 2020YFC2201504, by the Projects No. 11875062, No. 11947302, No. 12047503, and No. 12275333 supported by the National Natural Science Foundation of China, by the Key Research Program of the Chinese Academy of Sciences, Grant No. XDPB15, by the Scientific Instrument Developing Project of the Chinese Academy of Sciences, Grant No. YJKYYQ20190049, by the International Partnership Program of Chinese Academy of Sciences for Grand Challenges, Grant No. 112311KYSB20210012, and by DOE grant DE-FG02-13ER42020 (DVN).



\end{document}